\documentclass[twocolumn,showpacs,preprintnumbers,amsmath,amssymb]{revtex4}
\usepackage{graphicx,psfrag}
\usepackage{dcolumn}
\usepackage{bm}

\newcommand{\im}{{\rm i}}

\begin{document}


\title{Sigma meson in QCD sum rules using a two quark current with derivatives}

\author{Su Houng Lee}
 \email{suhoung@phya.yonsei.ac.kr}
\author{Kenji Morita}
 \email{morita@phya.yonsei.ac.kr}
\author{Kazuaki Ohnishi}
 \email{kohnishi@phya.yonsei.ac.kr}

\affiliation{Institute of Physics and Applied Physics, Yonsei University,
Seoul 120-749, Korea}
\date{\today}

\begin{abstract}
We study the $\sigma$ meson in QCD sum rules using a two quark interpolating field with derivatives.   In the constituent quark model, the $\sigma$ meson is composed of a quark and an antiquark in the relative p-wave state and is thus expected to have a larger overlap with an interpolating field that measures the derivative of the relative quark wave function.    While the sum rule with a current without derivative gives a pole mass of around 1 GeV, the present sum rule with derivative current gives a mass of around 550 MeV and a width of 400 MeV, that could be identified  with the $\sigma$ meson.
\end{abstract}


\maketitle

The existence of the light scalar meson or $\sigma(600)$ had been controversial
for a long time despite of its important role in chiral
dynamics in QCD physics \cite{van Beveren:1986ea, Kunihiro:2007yw}.
It is only recently that the existence has been confirmed
\cite{Hagiwara:2002fs,Eidelman:2004wy}, through the careful reanalyses of
the $\pi$-$\pi$ scattering phase shift\cite{Colangelo}, and the findings of the $\sigma$ pole
in the heavy particle decays such
as $D\rightarrow\pi\pi\pi$ and $\Upsilon(3S)\rightarrow\Upsilon\pi\pi$
\cite{Aitala:2000xu, Ishida:2001pt, Bugg:2003kj}.

On the other hand, the physical content of the
$\sigma$ meson is still controversial \cite{Close:2002zu}.
In addition to the usual picture of a $q\bar{q}$ state,
there are several plausible candidates like a four quark state $qq\bar{q}\bar{q}$
\cite{Jaffe:1976ig,Hooft}
and a $\pi\pi$ molecule, which can be mixed with a glueball. Also, the $\sigma$
meson may be a collective $q\bar{q}$ states just as the $\pi$ is, due to the
chiral symmetry \cite{Hatsuda:1985ey}.  Hence, it is important to confirm the existence of the $\sigma$ meson, and to investigate its nature on the basis of QCD.

In lattice QCD \cite{DeTar:1987xb},  most
of the calculations using a two quark interpolating field seem to
predict the ground state mass of scalar particle in the isospin 1
channel to be above 1.3 GeV
\cite{Bardeen04,Prelovsek04,Suganuma05,Mathur06,Burch06}, while
some predict it to be around 1 GeV \cite{McNeile06,Hashimoto}.  In
contrast, most lattice calculations based on a four quark
interpolating field consistently predict the mass to be
around 1 GeV for the $f_0,a_0$ \cite{Suganuma05,Ishii06} and around 600
MeV for the $\sigma$ \cite{Mathur06}.  The existence of the $\sigma$ with a two quark interpolating field was confirmed in a full QCD
simulation for the first time in Ref.\ \cite{Kunihiro:2003yj}.
It was argued that the disconnected diagram is crucially
responsible for reducing the $\sigma$ meson mass to that comparable to the $\rho$ meson mass for the lightest current quark mass achieved in the simulation. It is expected that if extrapolated to the chiral limit, the $\sigma$ meson mass would be such as $m_{\pi}<m_{\sigma}<m_{\rho}$.   On the other hand, for the $\kappa$, where no disconnected diagram contributes when a two quark interpolating field is used, the calculated mass was about two times that of the $K^*$ \cite{Wada07}.  These results suggest that the $\sigma$ might have a very different quark structure than the rest of the scalar nonet.

In this paper, we will study the $\sigma$ meson in QCD sum rules, which is another first principle approach to QCD \cite{Shifman:1978bx}.   The first QCD sum rule attempt was
made in Ref.\ \cite{Reinders:1981ww} with the usual two quark interpolating field
$\sim\bar{u}u(x)+\bar{d}d(x)$; a local operator.  Assuming the decay width to be zero, the calculated mass was found to be degenerate with its isospin 1 partner and around $m_{\sigma}=1.00\pm0.03$ GeV, which
is too large to be identify as the $\sigma$ meson.  On the other hand, recent works using four quark interpolating fields
\cite{Brito04,HJLee06,Matheus07,Chen07,Kojo08} seem to give a mass closer to that of the physical $\sigma$.

Our idea in this work is to observe that in a  non-relativistic quark model,
the quark and anti-quark in the $\sigma$ meson are in the relative p-wave state,
as can be deduced from its quantum number $J^{PC}=0^{++}$. The relative wave-function
has a node and a finite slope at the origin. Therefore, it is expected that the
$\bar{q}q$ operator with additional covariant derivative would have a larger
overlap with the physical $\sigma$ state
\cite{Chernyak:1981zz, Lee:1996au,Kim:1996cd}.   Thus we employ a two quark interpolating with derivatives, which will be called the non-local operator,
\begin{equation}
\mathcal{O}_{n}(x) \equiv
\frac{1}{\sqrt{2}}
\left\{
\bar{u}\left(\im z\cdot  \overleftrightarrow{D}\right)^{n}u(x)
+\bar{d}\left(\im z\cdot \overleftrightarrow{D}\right)^{n}d(x)
\right\}. \label{nonlocal}
\end{equation}
Here, $\overleftrightarrow{\im D}=\im D+(\im D)^\dag$,  $D_{\mu}=\partial_{\mu}-\im gA_{\mu}$ is the
covariant derivative and $z_{\mu}$ is a four-vector such that $z^{2}=-1$ in
order to probe the wave-function in the space-like direction. We note that for $n=0$, it corresponds to the local operator.
Of course, the quark operator in the interpolating field is that of a relativistic current quark.  Therefore, the current can still couple to the scalar meson that in the non relativistic limit is a p-wave state.  Nevertheless, we hope to increase the overlap with the physical $\sigma$ by employing an interpolating field that has the expected non relativistic limit.  Such interpolating currents with derivatives were previously used successfully to investigate the p-wave nucleon excited states
\cite{Lee:1996au,Kim:1996cd}.   This choice of interpolating field would  also be a natural choice in the lattice calculation, where the quark masses are still much larger than the physical limit.

Let us start by considering the following time-ordered two-point correlation
function
\begin{equation}
T_{n0}(q^{2}, z\cdot q, z^{2}) =
\im \int {\rm d}^{4}x e^{\im q\cdot x}
\langle 0 | {\rm T} \mathcal{O}_{n}(x) \mathcal{O}_{0}(0) | 0 \rangle.
\end{equation}
The correlation function will couple to the scalar meson when $n$ is even.  If $n$ is too large, the correlation will be dominated by the continuum contribution.  Therefore, the optimal choice for our analysis is $n=2$.
For $n=2$, the operator product expansion up to dimension 6 is calculated to be
\begin{align}
&T_{20}(q^{2}, z\cdot q, z^{2})
\nonumber\\
=&z^{2}
\left[
\frac{1}{8\pi^2} q^{4} \ln\left(-\frac{q^2}{\mu^2}\right)
+\frac{1}{4} \ln\left(-\frac{q^2}{\mu^2}\right)
\left\langle\frac{\alpha_{\rm s}}{\pi}G^{2}\right\rangle
\right.
\nonumber\\
&\left.
-\frac{16\pi}{81}\frac{1}{q^2}
\langle\sqrt{\alpha_{\rm s}}\bar{u}u\rangle^{2}
\right]
\nonumber\\
+&(z\cdot q)^{2}
\left[
-\frac{1}{8\pi^2} q^{2} \ln\left(-\frac{q^2}{\mu^2}\right)
-\frac{5}{24}\frac{1}{q^2}
\left\langle\frac{\alpha_{\rm s}}{\pi}G^{2}\right\rangle
\right.
\nonumber\\
&\left.
-\frac{272\pi}{81}\frac{1}{q^4}
\langle\sqrt{\alpha_{\rm s}}\bar{u}u\rangle^{2}
\right].
\end{align}
Note that we have calculated the OPE  up to dimension 6 operator.  It is important to calculate the OPE up to the so called quark tree diagram contributions when calculating the OPE of current-current correlation function.  This is so because purely gluonic operators are suppressed by loop effect that are typically suppressed by $1/4 \pi$, and higher order quark operators are suppressed by the  additional factors of $\alpha_s$.  This means that when calculating meson correlation function with multiple quark currents, the OPE should be investigated up to quark operators with quark number equal to 2 times that appearing in the current.  In our case, since our current involves two quark operators, the minimal quark operator that has to be considered is the four quark operator of dimension 6.

Picking up terms proportional to $z^2$,  subtracting out the continuum contribution that start from the threshold $s_0$, and then performing the
Borel transformation, leads to the following expression.
\begin{align}
 \Pi_0(M^2) &= -\frac{1}{4\pi^2}M^6 E_2(s_0/M^2) -
\frac{1}{4}M^2 E_0(s_0/M^2)G_0
\nonumber\\
&+ \frac{16\pi}{81}\langle \sqrt{\alpha_{\text{s}}}\bar{q}q
  \rangle^2, \label{lhs}
\end{align}
where $M$ is the Borel mass, and
\begin{equation}
 E_n(x) = 1-e^{-x}\sum_{m=0}^{n}\frac{x^m}{m!}.
\end{equation}
The theoretical side or the total OPE\cite{Shifman:1978bx}, is defined as Eq.(\ref{lhs}) with $s_0 \rightarrow \infty$.
The sum rule for the $\sigma$ is obtained by equating Eq.(\ref{lhs}) to the spectral density below the threshold $s_0$, which is dominated by the $\sigma$, through the Borel transformed dispersion relation.
\begin{equation}
 \Pi_0(M^2) = \frac{1}{\pi} \int_{0}^{\infty}\! {\rm d}s \, e^{-s/M^2}
  \text{Im}\Pi_0^{\sigma}(s). \label{dis-rel}
\end{equation}
For the spectral density coming from the $\sigma$, we include
a finite width by employing a simple Breit-Wigner form.
Namely,
\begin{equation}
 \text{Im}\Pi_0^{\sigma} = \frac{f\Gamma \sqrt{s}}{(s-m_\sigma^2)^2 + s\Gamma^2}.
\label{eq:BW}
\end{equation}
Here, $f$ is the overlap constant of the current with the physical $\sigma$.
The right hand side of Eq.(\ref{dis-rel}) with Eq.(\ref{eq:BW}) will be denoted as $\Pi_0^{\sigma}(M^2)$.

The QCD sum rule for the width and mass of the $\sigma$ is obtained by taking the ratio of the Eq.(\ref{dis-rel}) with its derivative with respect to the Borel mass.



\begin{equation}
 M^2
  \frac{\Pi_1^{\sigma}(M^2;m_\sigma,\Gamma)}{\Pi_0^{\sigma}(M^2;m_\sigma,\Gamma)} = M^2 \frac{\Pi_1(M^2;s_0)}{\Pi_0(M^2;s_0)},
\label{eq:sumrule}
\end{equation}
where
\begin{align}
 \Pi_1(M^2;s_0)&\equiv M^2 \frac{{\rm d}\Pi_0}{{\rm d}M^2}
\nonumber\\
 &= -\frac{3}{4\pi^2}M^6 E_3(s_0/M^2) + \frac{1}{4}M^2 E_1 G_0,
\end{align}
and $\Pi_1^\sigma(M^2) = M^2 \frac{{\rm d}\Pi_0^\sigma(M^2)}{{\rm d}M^2}$.
The left-hand side of Eq.(\ref{eq:sumrule}) becomes $m_\sigma^2$ in the narrow width
approximation.  In this limit, $m_\sigma^2$ is first plotted as a function of $M^2$ for different values of the threshold $s_0$.  From the plot with a threshold that gives the most stable plateau, the mass is determined from the value at the stable plateau region.  We remind the reader that such analysis has to be taken with the usual cautions as in any QCD sum rules involving ratios of derivatives\cite{Leinweber}.

In the present analysis, we will determine not only the mass but also the
width, with the following prescription:
\begin{enumerate}
 \item Fix a continuum threshold $s_0$, typically $0.8-2.0$ GeV$^2$.
 \item For each $s_0$, calculate the Borel window $M_{\text{min}}^2$ and
       $M_{\text{max}}^2$ by following the criterion that the power corrections and continuum correction should be respectively smaller then the theoretical side\cite{Shifman:1978bx} by certain fraction chosen as,
       \begin{align}
	M_{\text{min}}^2 &: \frac{\text{Power correction}}{\text{Total OPE
	sum}} \leq 0.3 -0.4\\
	M_{\text{max}}^2 &: \frac{\text{Pole contribution}}{\text{Total OPE sum
	}} \geq 0.3 -0.5
       \end{align}

 \item Compute the mass $m_\sigma$ by solving Eq.~\eqref{eq:sumrule}
       numerically for each fixed $\Gamma$, which are varied from 0 to
       500 MeV.
 \item Plot $m_\sigma$ as a function of $M^2$ for each $\Gamma$, and determine  $m_\sigma$ and $\Gamma$ that gives the most stable curve
       within the Borel window.
\end{enumerate}

\begin{figure}[tbh]
\begin{center}
\includegraphics[height=2.2in,keepaspectratio,angle=0]{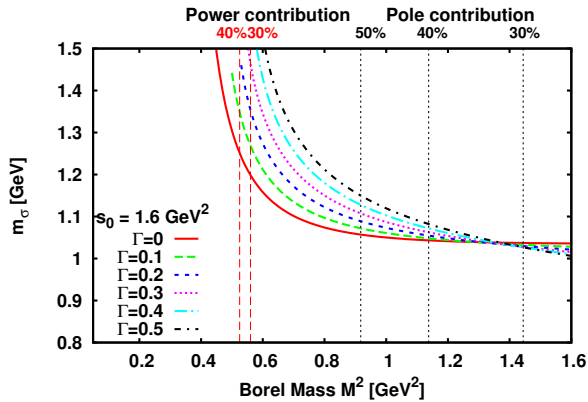}
\end{center}
\caption{Local operator: $m_{\sigma}$ as a function of $M^2$ for various
widths with $s_0=1.6$ ${\rm GeV}^2$. Vertical dotted lines denote the
 Borel masses at which continuum contribution becomes 30\%--50\% and
 power contribution becomes 30\%--40\%, respectively. }
\label{local_op1600}
\end{figure}

Now let us discuss numerical results. We begin with the $n=0$ case corresponding
to the local operator \cite{Reinders:1981ww}. In Fig.\ \ref{local_op1600},
we plot $m_{\sigma}$ as a function of $M^2$ for various widths with
$s_0=1.6$ ${\rm GeV}^2$.
We see that with $\Gamma=0$, we can reproduce the result of
Ref.\ \cite{Reinders:1981ww}, which shows a good Borel stability and gives the
mass of about 1 GeV. However, if we allow for a finite width, the stability is lost and we can not conclude that there is a resonance with that width.  Therefore, it is quite unlikely that the physical  $\sigma$ meson has strong coupling to the correlation function.  Instead, the stable Borel curve with small width seems to suggest a stronger coupling to $f_0(980)$ or $a_0(980)$, which are degenerate in the present calculation.

\begin{figure}[tbh]
\begin{center}
\includegraphics[height=2.2in,keepaspectratio,angle=0]{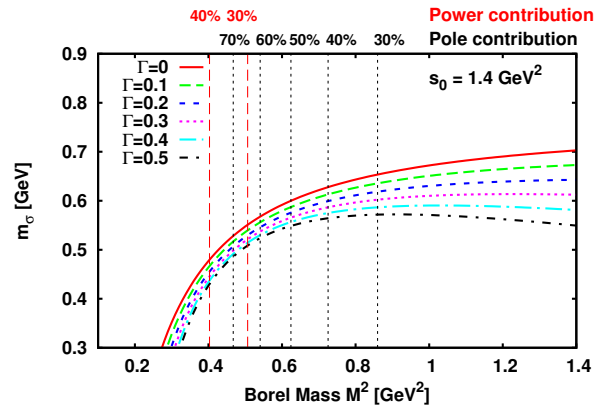}
\end{center}
\caption{Non-local operator: $m_{\sigma}$ as a function of $M^2$ for various
widths with $s_0=1.4$ ${\rm GeV}^2$.}
\label{nonlocal_op1400}
\end{figure}

For the non-local operator case with $n=2$, numerical analyses show that
we can find a stable plateau with a maximum in the Borel curves within the Borel window for a wide
range of the threshold parameter $s_0$.  The obtained mass and width are
$m_\sigma \simeq 350-950$ MeV and $\Gamma \sim 400-500$ MeV for
$s_0=1.2-2.0$ GeV$^2$. In Fig.\ \ref{nonlocal_op1400}, we show the result for
$s_0=1.4$ GeV$^2$.   We find that the Borel window is most stable and wide with the maximum within the window for $s_0=1.4 \pm 0.1 $ GeV$^2$.    It is very important to have a stable region or the extremum point within the Borel window.  Otherwise, the sum rule loses any predictive power and suggests that it is dominated by the continuum.  Even when $s_0=1.4$ GeV$^2$, we can see that such  a stable Borel curve with a maximum appears when we introduce  a large width. This is in contrast to the local
operator case, and suggest that the the correlation function is saturated with a resonance with a large width. We note that the maximum Borel mass is determined from the condition that the continuum contribution is less than 70 \%.  Also, in this case, contribution of the highest dimensional power
correction $(|\langle \bar{q}q \rangle|^2)$ is always less than 10\%
within the Borel mass region shown here.

Usually, meson sum rules are insensitive to the width.  An example being the vector meson sum rule; while the OPE of the $\rho$ and $\omega$ sum rule are identical their widths are very different.  On the other hand, our sum rules are of higher dimensions than the vector meson sum rule. This means that the continuum contributes with higher moments and that the sum rules are more sensitive to the continuum values.  This also means that the sum rules are more sensitive to the width. If the sum rule were of too high dimension, then the sum rule would be dominated by the continuum and we could not obtain any information about the pole or the width.  What we have shown is that for the sum rules at hand, the dimensions are optimal such that the  sum rules are sensitive to the width but at the same time still dominated by the pole, especially for the sum rule with non-local current.  

To further check the consistency in our approach, it would be useful to construct the three-point function among the two-quark current with derivative and the two pseudo-scalar currents, which will couple to the sigma and the pions, respectively.  The analysis of such a function will provide information for the $\sigma$-$\pi$-$\pi$ coupling, from which one could directly calculate the sigma width and check the consistency with the present calculation. This will be left as a future work.
Another related work would be to introduce an effective diquark scalar
field in QCD \cite{Grinstein}, construct the current for the sigma meson using the
diquark current, and then perform a similar QCD sum rule
analysis. The lowest dimensional current will not include a
derivative as two quarks and two anti-quarks in $s$-wave states
will form a scalar meson in the constituent quark model. The
QCD sum rule analysis will be more effective in concentrating on the attraction between the scalar diquarks and thus  more effective than the previous QCD sum rule analysis
using the four quark interpolating field.

To summarize, we have investigated the $\sigma$ meson in QCD sum rules by
using the non-local $\bar{q}q$ operator with a covariant derivative. Such a choice of current was motivated by the fact that the wave-function is a p-wave in a naive non-relativistic quark model.   We have confirmed that with such a choice of current, the sum rule is consistent with a strong coupling to the $\sigma$ meson. We note that it is necessary to take into account both the mass and the width
simultaneously: If we set the width to zero, we do not obtain the Borel
stability at all. The best estimate that gives the most stable plateau is with a  mass $m_{\sigma}= 550$ MeV and $\Gamma_{\sigma}= 400$ MeV in the chiral limit, which is the first
theoretical prediction based on a first principle QCD calculation.  Our work also confirms that the $\sigma$ has a large $\bar{q}q$ component.
Therefore, it would be interesting to try similar approach in lattice calculation.
There are several things to improve in the future. First, in the present analysis,
the $I=0$ channel ($\sigma$) and the $I=1$ channel ($\delta$) are degenerate.
To lift the degeneracy, we have to introduce higher dimensional operators or effectively take into account such contributions through the instanton effect
\cite{Dmitrasinovic:1996fi, Naito:2003zj, Elias:1998bq}.   This is so because  single instanton configurations are known to contribute to correlation function of scalar currents\cite{Shuryak93}.  Moreover, $a_0(980)$ is expected to have a large tetraquark component as has been shown in recent QCD sum rule calculations\cite{Brito04,HJLee06,Matheus07,Chen07,Kojo08}.
Secondly, it is necessary to take into account the $\alpha_s$ correction to the perturbative contribution and the corrections from the
finite current quark masses.  Through such extension, we can investigate the whole scalar nonet,  among which the $\kappa$(800) meson is of great interest
\cite{Bugg:2003kj, Aitala:2002kr}. Third, it would be interesting
to study the other p-wave mesons systematically by using the non-local operator,
as in Ref.\ \cite{Reinders:1981ww}.

\begin{acknowledgments}
We would like to thank APCTP for sponsoring the workshop on 'Hadron Physics
at RHIC' and the organizers of NFQCD2008 at YITP.  The discussions during these workshops have lead to the final version of this paper.
We are particularly grateful to Teiji Kunihiro for useful discussions.
\end{acknowledgments}


\end{document}